\documentclass[fleqn,12pt,twoside]{article}
\usepackage{espcrc1}

\usepackage{graphicx}

\newcommand{\AmS}{{\protect\the\textfont2
  A\kern-.1667em\lower.5ex\hbox{M}\kern-.125emS}}

\title{Results on Photon Production in Au+Au Collisions at RHIC}

\author{Klaus Reygers\address[MS]{University of M\"{u}nster, 
        D-48149 M\"{u}nster, Germany}
        for the PHENIX Collaboration\footnote{for the full PHENIX 
        Collaboration author list and acknowledgements, see Appendix 
        "Collaborations" of this volume.}}

\begin{document}

\maketitle

\begin{abstract}
The status of the search for direct photons in Au+Au collisions 
at $\sqrt{s_{NN}} = 130$~GeV and  $\sqrt{s_{NN}} = 200$~GeV 
with the PHENIX experiment is presented. Within errors, no excess 
of direct photons was found in a first analysis pass done on a 
limited data set. Significantly reduced systematic and statistical 
uncertainties are expected in future analyses.
\end{abstract}

\section{Introduction}
\label{sec:intro}
The aim of this analysis is to search for direct photons, i.e. 
photons that are not produced in hadron decays such as 
$\pi^0 \rightarrow \gamma \gamma$ or $\eta \rightarrow \gamma \gamma$.
Direct photons are generally subdivided into prompt photons from initial 
hard parton scatterings and thermal photons from a possible 
quark-gluon plasma and from the hadron gas. 

Prompt photons are produced before a possible quark-gluon plasma
has formed. Above a transverse momentum of $p_T \approx 3-5$~GeV 
prompt photons are expected to be the dominant source of direct photons
in Au+Au collisions at RHIC. In p+p reactions direct photons are used 
to obtain information about the gluon distribution in the proton since 
their production in p+p is dominated by quark-gluon Compton 
scattering ($g q \rightarrow \gamma q$). 
By measuring direct photons in Au+Au at RHIC one is sensitive to 
modifications of the parton distribution functions in the nucleus 
({\it nuclear shadowing}) \cite{shadowing}. 

The agreement between perturbative QCD calculations and 
experimental direct photon data in p+p and p+A can be improved 
if a momentum component $k_T$ of the initial partons transverse to 
the beam axis is introduced in the calculations \cite{kt}. 
In nucleus-nucleus collisions one expects a stronger $k_T$ than 
in p+p due to e.g. multiple soft scatterings of the incoming nucleons.
As demonstrated in \cite{dumitru} the magnitude of this intrinsic $k_T$ can 
be constrained by measurement of direct photons in nuclear collisions. 

Thermal photons are produced over almost the entire evolution of a
nucleus-nucleus collision: in a possible quark-gluon plasma phase and in 
the later hadron gas phase. The strongest contribution from 
thermal photons is expected roughly in the range $p_T \approx 1 - 5$~GeV.
Theoretically, the entire hydrodynamical
evolution of a nucleus-nucleus reaction needs to be modeled to describe
thermal photon production. A comparison of these models with data
constrains the initial temperature of the reaction system. 
In order to obtain information about the existence of a 
quark-gluon plasma one needs to compare models with and without 
phase transition to the data \cite{peitzmann}.

The first observation of direct photons in heavy ion collisions 
was achieved at the CERN SPS by the WA98 experiment \cite{wa98_photons}. 
At RHIC, a substantial suppression of $\pi^0$ production at high $p_T$
in central reactions was found that was not present at SPS energies
\cite{phenix_sup}. The reduced background from neutral pions 
could result in a clearer direct photon signal at RHIC. 
Moreover, the passage of quark jets through the excited nuclear
medium might be an important new source of direct photons at RHIC, 
such that the subdivision of direct photons into prompt and thermal 
photons would have to be extended \cite{fries}.

\section{Data Analysis}
\label{sec:analysis}
Direct photon measurements (even in p+p) are generally considered to be 
very challenging. It is therefore an advantage of the PHENIX 
experiment that photons can be measured in different ways:
directly with an electromagnetic calorimeter and also via their conversions 
in $e^+e^-$-pairs. Here we focus on the photon measurement 
with the electromagnetic calorimeter (EMCal) of the PHENIX experiment.
The biggest experimental uncertainties in photon measurements with 
calorimeters are generally background from hadrons that are misidentified 
as photons and errors of the energy scale of the detector. In heavy ion 
collisions an additional uncertainty results from the high 
multiplicity environment leading to overlapping showers in the 
detector.
\begin{figure}
   \centerline{\includegraphics[height=5.0cm]{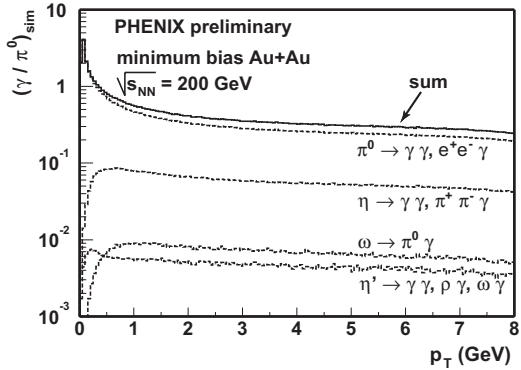}}
   \caption{Simulation result for the $\gamma$/$\pi^0$ ratio in 
            $\sqrt{s_{NN}} = 200$~GeV Au+Au collisions expected from 
            hadronic decays.}
   \label{fig:gam_pi0_ratio}
\end{figure}

In p+p reactions direct photons at sufficiently high $p_T$ 
are usually identified on an event-by-event basis with 
the help of certain isolation cuts. This is more difficult in 
heavy ion reactions and therefore the approach in this analysis is to 
try to find direct photons on a statistical basis. The inclusive photons 
(hadron decay photons and direct photons) are measured and then 
the background of photons from hadron decays is subtracted.

The PHENIX EMCal consists of two sub-detectors, a lead-scintillator
calorimeter (PbSc, 6 sectors) and a lead-glass calorimeter 
(PbGl, 2 sectors). The lead-glass detector was previously used in the WA98 
experiment. Each sector covers a pseudorapidity range of $|\eta| < 0.35$ and
an azimuthal angle of $\phi \approx 22.5^{\circ}$. 
Both sub-detectors are highly segmented 
($\Delta \phi \times \Delta \eta \approx 0.01 \times 0.01$) such that
the two decay photons of a $\pi^0$ are well separated up 
to neutral pion momenta of $p_T \approx 20$~GeV.    
The different detection mechanisms of the two sub-detectors
(measurement of scintillation light in PbSc and detection of Cherenkov
photons in PbGl) result in a different response to hadrons.
Thus, PbSc and PbGl provide photon measurements with different systematics. 

The extraction of the inclusive photon spectra starts by identifying 
photon-like hits in the EMCal. Hadrons on the average produce showers
with larger lateral extensions than photons. Thus, cuts on the shower shape 
are used to identify photons both in PbSc and PbGl. In addition, the 
time-of-flight of the showers is used as an identification criterion.
The next analysis step is the subtraction of background from 
charged hadrons, neutrons, anti-neutrons, and particles not coming 
from the vertex. This is done statistically and not on an event-by-event
basis. In order to determine the background of charged hadrons in the 
sample of photon-like hits, identified tracks from the PHENIX tracking
system are projected to the EMCal surface. Random associations between
photons and charged tracks are corrected for by an event-mixing technique.
Above photon transverse momenta of $p_T \approx 1.5$~GeV the background of 
charged hits in the photon-like sample is found to be less than 10\%.
The background of neutrons and anti-neutrons in the sample of neutral 
EMCal hits, that is obtained after subtracting the charged background,
has to be determined from simulation. The assumed input spectra
of neutrons and anti-neutrons needed in the simulation are estimated 
based on measured $p$ and $\bar{p}$ spectra. The contribution of 
neutrons and anti-neutrons to all neutral EMCal hits is found to be less
than 6\%. In the last analysis step a photon efficiency correction
is applied which takes the energy resolution of the detector, shower 
overlaps and photon losses due to identification cuts into account.

\begin{figure}
   \centerline{\includegraphics[height=5.0cm]{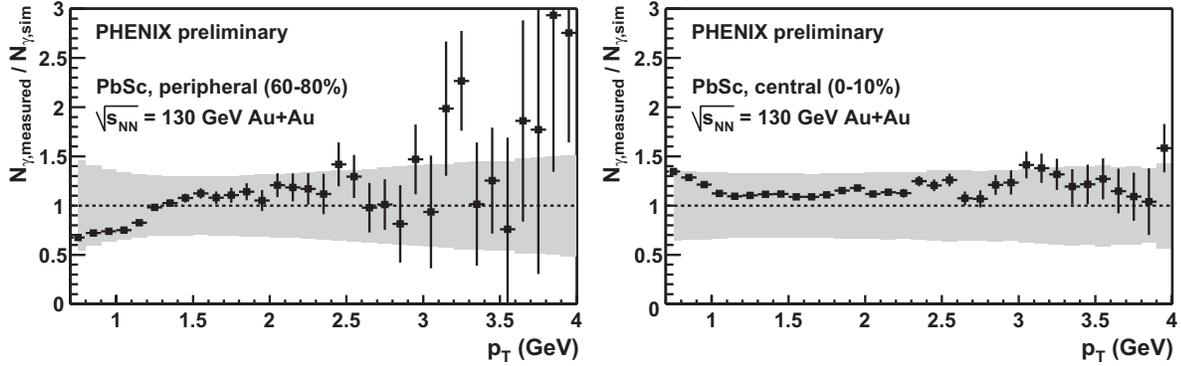}}
   \caption{Ratio of measured photons to expected background photons from
            hadron decays in Au+Au at $\sqrt{s_{NN}} = 130$~GeV.
            The centrality classes are specified by fractions 
            of the total geometrical Au+Au cross section.
            The shaded band indicates the systematic error.}
   \label{fig:gamma_ratio_130}
\end{figure}

The expected photons from hadron decays are determined in a simulation 
that takes the measured neutral pion spectrum as input. Exactly
the same data set is used for the neutral pion and the photon
analysis. The neutral pion spectra obtained in this analysis agree
with those presented in \cite{enterria} within the current 
systematic error of 25\%.
An $\eta$-spectrum has not yet been measured. Therefore $m_T$-scaling
is assumed for the $\eta$ and other hadrons. The result of the
hadron decay simulation is shown in Figure~\ref{fig:gam_pi0_ratio}.

\section{Results}
\label{sec:results}
\begin{figure}
   \centerline{\includegraphics[height=5.0cm]{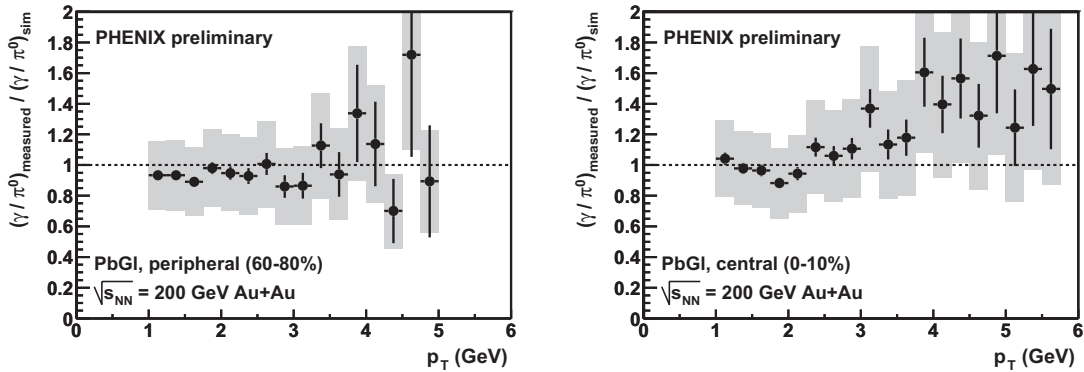}}
   \caption{Measured over expected $\gamma/\pi^0$ ratio in Au+Au 
            collisions at $\sqrt{s_{NN}} = 200$~GeV. The shaded boxes
            represent the estimated 1$\sigma$ absolute systematic 
            errors of the data points.}
   \label{fig:gamma_ratio_200}
\end{figure}

In order to search for non-hadronic photon sources the ratio 
of all measured photons to the expected background photons 
from hadron decays is shown as a function of $p_T$ for 
Au+Au collisions at $\sqrt{s_{NN}} = 130$~GeV 
(Figure~\ref{fig:gamma_ratio_130}) and 200 GeV 
(Figure~\ref{fig:gamma_ratio_200}).

The results in Figure~\ref{fig:gamma_ratio_130} are from an analysis  
that is similar to the one described in the previous section, but 
the charged background was determined from simulation and not from real 
data. The statistics in the 130~GeV data is small 
compared to the 200~GeV data such that the shape of the $\pi^0$ spectrum
is not as well constrained as in the 200~GeV analysis. 
In Figure~\ref{fig:gamma_ratio_200} the results for 
Au+Au at $\sqrt{s_{NN}} = 200$~GeV are presented in terms
of the ratio of the measured $\gamma/\pi^0$ ratio to the 
simulated $\gamma/\pi^0$ ratio. In the $\gamma/\pi^0$ ratio 
systematic errors partially cancel. The current systematic errors 
are dominated by uncertainties of the photon and neutral pion 
reconstruction efficiencies. 

As can been seen from Figures~\ref{fig:gamma_ratio_130} and
\ref{fig:gamma_ratio_200} no direct photon signal is observed 
within current errors both in Au+Au collisions 
at $\sqrt{s_{NN}} = 130$~GeV and $\sqrt{s_{NN}} = 200$~GeV. 
The systematic uncertainties are expected to be reduced significantly
in the future with further analysis.

\end{document}